\documentclass[11pt,twoside]{article}
\usepackage[pdftex]{graphicx}
\usepackage{amsmath}
\usepackage{amssymb}
\usepackage{cite}

\begin{document}
\newcommand{\pst}{\hspace*{1.5em}}

\newcommand{\rigmark}{\em Journal of Russian Laser Research}
\newcommand{\lemark}{\em Volume 30, Number 5, 2009}

\newcommand{\be}{\begin{equation}}
\newcommand{\ee}{\end{equation}}
\newcommand{\bm}{\boldmath}
\newcommand{\ds}{\displaystyle}
\newcommand{\bea}{\begin{eqnarray}}
\newcommand{\eea}{\end{eqnarray}}
\newcommand{\ba}{\begin{array}}
\newcommand{\ea}{\end{array}}
\newcommand{\arcsinh}{\mathop{\rm arcsinh}\nolimits}
\newcommand{\arctanh}{\mathop{\rm arctanh}\nolimits}
\newcommand{\bc}{\begin{center}}
\newcommand{\ec}{\end{center}}

\thispagestyle{plain}

\label{sh}


\begin{center} {\Large \bf
\begin{tabular}{c}
SUPERPOSITION  PRINCIPLE FOR QUBIT STATES 
\\[-1mm]
IN SPIN-PROJECTION MEAN REPRESENTATION

\end{tabular}
 } \end{center}

\bigskip

\bigskip

\begin{center} {\bf
Fedorov A.Yu.$^{1,\,2,\,3}$ and Man'ko V.I.$^{1,\,2,\,3}$
}\end{center}

\medskip

\begin{center}
{\it
$^1$Lebedev Physical Institute, Russian Academy of Sciences\\
Leninskii Prospect 53, Moscow 119991, Russia

\smallskip

$^2$Moscow Institute of Physics and Technology (State University)\\
Institutskii per. 9, Dolgoprudnyi, Moscow Region 141700, Russia

\smallskip

$^3$Russian Quantum Center, Skolkovo, Moscow 143025, Russia}

\smallskip

$^*$Corresponding author e-mail:~~~manko~@~sci.lebedev.ru\\
\end{center}

\begin{abstract}\noindent
The superposition principle of two qubit states is formulated as nonlinear addition rule of mean spin-projection onto three perpendicular directions. The explicit expression for the mean value determining the superposition state in term of the two initial pure states of the qubit is obtained. Possibility to check the superposition expression in experiments is discussed. 
\end{abstract}

\medskip


\section{Introduction}
\pst
In quantum mechanics the states are defined by wave functions satisfying linear Schrodinger equation [Schrodinger 1926][1,2] or density matrix introduced by Landau [Landau,Lifschitz] [3],see also [Dirac] [4],[Davidov] [5]. There are other representations of quantum states, for example, the Wigner function was introduced by Wigner [6]. Also, for example, there is a representation of states through the Husimi functions [7]. In 1963, Glaubeh [8] and Sadarshan [9] introduced their presentation - the representation of Glauber-Sudarshan [10]. In 1997, a probabilistic representation of quantum states was introduced [11], which forms the basis of this work. In this representation, the state is given by probability distributions — quantum tomography, which can be measured experimentally [12].In this paper, we consider a new representation related to the system — spin-$\frac{1}{2}$, which follows from the probabilistic representation. We note that a representation similar to the Wigner representation was constructed for spin-$\frac{1}{2}$ Stratonovich [13].

Since the Schrödinger equation is linear one the superposition of its solutions is also solution of the equation. This property corresponds to physical interference phenomenon.
We address in this paper the following problem. Is it possible to formulate the superposition principle using not the wave functions but using some statistical characteristics of physical observables e.g. mean values of the observables. These wave functions are expressed in terms of the mean values of the spin projection. Due to this superposition of wave functions gives the possibility to describe this superposition in terms of corresponding linear property of the mean values.

We consider this approach using the simple example of spin-$\frac{1}{2}$ system. The superposition of the states was expressed in probability representation of quantum states using probability distribution determining in this representation these states.
We present here another form of the states superposition’s expressed in terms of the spin observables mean values. This approach can be extended to other systems using the linear equations for tomographic probability distributions see e.g. [14].

In this paper, we show that the spin wave functions of particles with spin-$\frac{1}{2}$ can be expressed in terms of the mean values of the spin projection on the orthogonal coordinate axes. We also show what kind the density matrix ρ will have in these terms. On the basis of this representation, we obtain the superposition formula of these wave functions in terms of the mean values of the spin projections. The states, in our case, the spin state, can be expressed in terms of classical quantities that can be measured experimentally. Thanks to this view, it is assumed that the laws of quantum physics can be verified experimentally.

The paper is organized as follows.

In Sec. 2 we show that the spin wave functions of particles with spin-$\frac{1}{2}$ can be expressed in terms of the average values of the spin projection on the orthogonal coordinate axes. We also show what kind the density matrix ρ will have in these terms.
In Sec.~3 On the basis of this representation, we obtain the superposition formula of these wave functions in terms of the mean values of the spin projections. The states, in our case, the spin state, can be expressed in terms of classical quantities that can be measured experimentally. Thanks to this view, it is assumed that the laws of quantum physics can be verified experimentally.
In Sec.~4 
we review the results.

\section{Spin density matrix of a particle with spin s = $\frac{1}{2}$. The relationship of the components of the spin density matrix with the average values ​​of the projections of the spin on the orthogonal x, y, z axes, respectively.}
\pst
The states in quantum physics can be described using wave functions, these are the so-called pure states, also in a more general form using a density matrix. For pure states, the density matrix is expressed in terms of wave functions as: $\rho=|\psi\rangle\langle\psi|$. The spin density matrix of a particle with spin 1/2 is a second rank spinor $\rho^{\lambda^\mu}$, normalized by the condition:
$\rho_\lambda^\lambda=\rho_1^1+\rho_2^2=1$ and satisfying the “Hermitian” condition:$(\rho_\mu^\lambda)^*=\rho_\lambda^\mu$. The states of a quantum system determined by the vectors $|\psi\rangle$ in a Hilbert space have the property that the superpositions of the vectors $c_1|\psi_1\rangle+c_2|\psi_2\rangle$  with complex coefficients c1 and c2 always describe the other states of the system. The superposition of the vectors corresponds to superposition of wave functions $c_1\psi_1(q)+c_2\psi_2(q)$ describing the states of the systems.

To begin with, we derive how the mean values of the projections of the spin on the orthogonal axes (x,y,z) are related to the components of the spin density matrix. The diagonal components of the density matrix determine the probabilities of the values +1/2  and -1/2  spin projections of a spin 1/2  particle on an axis z. Therefore, the mean value of this projection:
\be
\overline{s}_z=\frac{1}{2}(\rho_1^1-\rho_2^2).
\ee

Given the normalization $\rho_1^1+\rho_2^2=1$ ,we obtain:
\be
\rho_1^1=\frac{1}{2}+\overline{s}_z,\ \ \rho_2^2=\frac{1}{2}-\overline{s}_z. 
\ee

In the pure state, the average value $s_\pm=s_x \pm is_y$ is calculated as:
\be
\overline{s}_+=\psi^{\lambda^*} \widehat{s}_+ \psi^\lambda,\ \ \overline{s}_-=\psi^{\lambda^*} \widehat{s}_- \psi^\lambda.
\ee

The spin vector for a spin-$\frac{1}{2}$ particle is expressed as follows, see [3]:
\be
\widehat{\boldsymbol{s}}\boldsymbol{=}\frac{1}{2}\widehat{\boldsymbol{\sigma }}
\ee

where $\widehat{\sigma }_x,\widehat{\sigma }_y,\widehat{\sigma }_z$ are the Pauli matrices:
\be
\widehat{\sigma }_x=\left( \begin{array}{cc}
0 & 1 \\ 
1 & 0 \end{array}
\right), \     \widehat{\sigma }_y=\left( \begin{array}{cc}
0 & -i \\ 
i & 0 \end{array}
\right),\   \widehat{\sigma }_z=\left( \begin{array}{cc}
1 & 0 \\ 
0 & -1 \end{array}
\right).                         
\ee

Then the operators $\widehat{s}_\pm $ are expressed by matrices:
\be
{\hat{s}}_+=\left( \begin{array}{cc}
0 & 1 \\ 
0 & 0 \end{array}
\right),   \ \  {\hat{s}}_-=\left( \begin{array}{cc}
0 & 0 \\ 
1 & 0 \end{array}
\right).
\ee

So, substituting these values in (4), we find:  
\be
\overline{s}_+={\psi }^{1*}{\psi }^2,\ \     \overline{s}_-={\psi }^{2*}{\psi }^1.
\ee

Accordingly, in a mixed state will be:
\be
{\ \rho }^1_2=\overline{s}_-,\ \ {\rho }^2_1=\overline{s}_+.
\ee

Thus, we associate the components of the spin matrix with the density of particles with spin s = $\frac{1}{2}$ with the mean values ​​of the spin projections on
orthogonal x, y, z axes, that is, through $\overline{s_z}$,$\ {\overline{s}}_x$,$\ {\overline{s}}_y$.

The aim of this work is to obtain an explicit form of the spin wave function for a particle with spin s = $\frac{1}{2}$ through the mean values of the spin projections onto the orthogonal x, y, z axes. The derivation of a formula for the superposition of two pure non-orthogonal states based on this representation.
A new idea is that the spin state of a particle can be expressed in terms of a probability distribution. Namely, in the works [15-18] it was found that the density matrix for a particle with spin s = $\frac{1}{2}$ is expressed in terms of three values $p_1, p_2, p_3$ as follows:
\be
\rho =\left( \begin{array}{cc}
p_3 & {(p}_1-\frac{1}{2}\mathrm{)}\mathrm{-}\mathrm{i(}p_2-\frac{1}{2}) \\ 
{(p}_1-\frac{1}{2}\mathrm{)\ }\mathrm{+i(}p_2-\frac{1}{2}) & 1-p_3 \end{array}
\right).
\ee

Where $p_1$,$\mathrm{\ }p_2,\ p_3$ are probabilities to have the spin +1/2  projection on the axes x,y,z, respectively.

$p_1$,$\mathrm{\ }p_2,\ p_3$ satisfying conditions:
\be
0\le p_1,\mathrm{\ }p_2,\ p_{3\ }\le 1.
\ee

Due to the properties of the density matrix, these values ​​are subject to additional restrictions:
\be
{{(p}_1-\frac{1}{2})}^2+{{(p}_2-\frac{1}{2})}^2+{{(p}_3-\frac{1}{2})}^2\le \frac{1}{4}.
\ee

Equality is achieved when considering pure states.

On the basis of this density matrix, we obtain the formula for the spin state of a particle with spin $s=\frac{1}{2}$ through the probability distributions $p_1, p_2, p_3$. Arbitrary Paul's spinor is expressed as follows, see [3]:
\be
\left|\mathrm{\chi }\right\rangle =\left(\  \begin{array}{c}
{\chi }_1 \\ 
{\chi }_2 \end{array}
\ \right).
\ee

Make a replacement ${\chi }_1=e^{ia}a,\  {\chi }_2=e^{i\beta }b$, where $a>0,b>0$. Then, substituting into formula (12), we get:
\be
\left|\mathrm{\chi }\right\rangle =e^{i\alpha }\left(\  \begin{array}{c}
a \\ 
be^{i\gamma } \end{array}
\ \right),\ \ \gamma =(\beta -\alpha ).
\ee

Then, using the formula $\rho=|\psi\rangle\langle\psi|$, we get:
\be
\rho =\left( \begin{array}{cc}
aa^* & ab^*e^{-i\gamma } \\ 
a^*be^{i\gamma } & bb^* \end{array}
\right).
\ee

Comparing with (9), we obtain that the spin wave function is expressed in terms of probability distributions $ p_1, p_2, p_3$ as follows:
\be
\left|\chi \right\rangle =e^{i\alpha }\left( \begin{array}{c}
\sqrt{p_3} \\ 
\sqrt{1-p_3}e^{i\gamma } \end{array}
\right).
\ee

where we introduce probabilities given the relationships:
\be
a^2=p_3,\ \ \ \ \ b^2=1-p_3,\ \ \ \ \ {\mathrm{cos} \gamma \ }=\frac{p_1-\frac{1}{2}}{\sqrt{p_{3\ }\left(1-p_{3\ }\right)}},\ \ \ \ \ {\mathrm{sin} \gamma \ }=\frac{p_2-\frac{1}{2}\ }{\sqrt{p_{3\ }\left(1-p_{3\ }\right)}}\ .
\ee

Phase $\alpha$  is an arbitrary angular parameter. The value of this parameter does not change the pure state density operator $\widehat{\rho }=\left|\chi \right\rangle \langle \chi |$, which is a gauge-invariant operator of the projector.
Further, using relations (2,8) and (9), we express the probability distributions $p_1, p_2, p_3$ in terms of the average values of the spin projection $\ {\overline{s}}_x$,$\ {\overline{s}}_y,{\overline{s}}_z$ on the orthogonal axes x, y, z. We get:
\be
p_3=\frac{1}{2}+{\overline{s}}_z, \ \     p_2=\frac{1}{2}+{\overline{s}}_y,\ \      p_1=\frac{1}{2}+{\overline{s}}_x.                              
\ee

Then formulas (9) and (15) are transformed through $\ {\overline{s}}_x$,$\ {\overline{s}}_y,{\overline{s}}_z$ as:
\be
\left|\chi \right\rangle =e^{i\alpha }\left( \begin{array}{c}
\sqrt{\frac{1}{2}+{\overline{s}}_z} \\ 
\sqrt{\frac{1}{2}-{\overline{s}}_z}\left(\frac{{\overline{s}}_x}{\sqrt{(\frac{1}{2}+{\overline{s}}_z)(\frac{1}{2}-{\overline{s}}_z)}}+i\frac{{\overline{s}}_y}{\sqrt{(\frac{1}{2}+{\overline{s}}_z)(\frac{1}{2}-{\overline{s}}_z)}}\right) \end{array}
\right).
\ee

\be
\rho =\left( \begin{array}{cc}
\frac{1}{2}+{\overline{s}}_z & {\overline{s}}_x\mathrm{-}\mathrm{i}{\overline{s}}_y \\ 
{\overline{s}}_x\mathrm{+i}{\overline{s}}_y & \frac{1}{2}-{\overline{s}}_z \end{array}
\right).
\ee

Thus, formula (18) is a general form of the spin wave function of a particle with spin 1/2 through the mean values of the spin projections $\ {\overline{s}}_x$,$\ {\overline{s}}_y,{\overline{s}}_z$. Formula (19) expresses the general form of the density matrix for a particle with spin $s = \frac{1}{2}$ in terms of the mean values of the spin projections $\ {\overline{s}}_x$,$\ {\overline{s}}_y,{\overline{s}}_z$, on the x, y, z axes, respectively.

\section{Derivation of the superposition formula for two pure non-orthogonal qubit states in the new representation.}
\pst

Consider the superposition of two pure non-orthogonal qubit states through the introduced new representation.
Based on formula (18), the first state $\left|{\psi }_1\right\rangle $  is defined as:
\be
\left|{\psi }_1\right\rangle =\left( \begin{array}{c}
\sqrt{\frac{1}{2}+{\overline{s}}_z} \\ 
\sqrt{\frac{1}{2}-{\overline{s}}_z}e^{i\gamma 1} \end{array}
\right).
\ee

Then, the bra vector has the form:                  
\be
\langle {\psi }_1|=\left(\sqrt{\frac{1}{2}+{\overline{s}}_z},\ \sqrt{\frac{1}{2}-{\overline{s}}_z}e^{-i\gamma 1}\right).                      \ee

The explicit density matrix $\rho =|{\psi }_1\rangle \langle {\psi }_1|$ of this pure state expressed in terms $\overline{s}_z$,$\ {\overline{s}}_x$,$\ {\overline{s}}_y$  reads:
\be
|{\psi }_1\rangle \langle {\psi }_1|=\left( \begin{array}{cc}
\frac{1}{2}+{\overline{s}}_z & \sqrt{\left(\frac{1}{2}+{\overline{s}}_z\right)\left(\frac{1}{2}-{\overline{s}}_z\right)}e^{-i\gamma 1} \\ 
\sqrt{\left(\frac{1}{2}+{\overline{s}}_z\right)\left(\frac{1}{2}-{\overline{s}}_z\right)}e^{i\gamma 1} & \frac{1}{2}-{\overline{s}}_z \end{array}\right).
\ee

Let us introduce the second state-vector of the qubit expressed in terms of the three mean values ${\overline{S}}_x,{\overline{S}}_y,{\overline{S}}_z,\ $of the spin-1/2 projections on the axes x,y,z, respectively, as:
\be
\left|{\psi }_2\right\rangle =\left( \begin{array}{c}
\sqrt{\frac{1}{2}+{\overline{S}}_z} \\ 
\sqrt{\frac{1}{2}-{\overline{S}}_z}e^{i\gamma 2} \end{array}
\right).                                              
\ee

In order to derive the general formula for the superposition of two pure nonorthogonal states of the qubit expressed in terms of spinors:
\be
\left|{\chi }_1\right\rangle =\left( \begin{array}{c}
a_1 \\ 
a_2 \end{array}
\right),\ \  \left|{\chi }_2\right\rangle =\left( \begin{array}{c}
b_1 \\ 
b_2 \end{array}
\right).
\ee

Such that ${\left|a_1\right|}^2+{\left|a_2\right|}^2={\left|b_1\right|}^2+{\left|b_2\right|}^2$=1 and defined as:
\be
\left|\chi \right\rangle =[c_1\left|{\chi }_1\right\rangle +c_2\left|{\chi }_2\right\rangle ].    
\ee

We consider the matrix:
\be
|\chi \rangle \langle \chi |=\left( \begin{array}{c}
{c_1a}_1+c_2b_1 \\ 
{c_1a}_2+c_2b_2 \end{array}
\right)\left({c^*_1a}^*_1+c^*_2b^*_1,{c^*_1a}^*_2+c^*_2b^*_2\right).            \ee

The obtained matrix provides the density matrix of the superposition state ${\rho }_{\chi }$ if one takes into account the normalization condition, i.e. T$=\langle \chi |\left|\chi \right\rangle $=Tr$\left|\chi \right\rangle \langle \chi |$ provides relation Tr${\rho }_{\chi }$=1. Thus the density matrix is defined as ${\rho }_{\chi }=\left|\chi \right\rangle \langle \chi |$ /$\langle \chi |\left|\chi \right\rangle $:
\be
{\rho }_{\chi }=\frac{1}{{\left|{c_1a}_1+c_2b_1\right|}^2+{\left|{c_1a}_2+c_2b_2\right|}^2}\left( \begin{array}{cc}
{\left|{c_1a}_1+c_2b_1\right|}^2 & \left({c_1a}_1+c_2b_1\right)\left({c^*_1a}^*_2+c^*_2b^*_2\right) \\ 
{{(c}^*_1a}^*_1+c^*_2b^*_1)({c_1a}_2+c_2b_2) & {\left|{c_1a}_2+c_2b_2\right|}^2 \end{array}
\right).
\ee

The superposition state with the density matrix ${\rho }_{\chi }$ can be expressed in terms of the three mean values ${\overline{\mathcal{S}}}_x,{\overline{\mathcal{S}}}_y,{\overline{\mathcal{S}}}_z\ $of the spin-1/2 projections on the axes x,y,z, respectively, then:
\be
{\rho }_{\chi }=\left( \begin{array}{cc}
{\overline{\mathcal{S}}}_z+\frac{1}{2} & {\overline{\mathcal{S}}}_x\mathrm{-}\mathrm{i}{\overline{\mathcal{S}}}_y\  \\ 
{\overline{\mathcal{S}}}_x\mathrm{\ }\mathrm{+i}{\overline{\mathcal{S}}}_y & \frac{1}{2}-{\overline{\mathcal{S}}}_z \end{array}
\right).                                         
\ee

This density matrix is associated with the normalized pure state $\left|{\chi }_0\right\rangle $ of the form:
\be
\left|{\chi }_0\right\rangle =\left( \begin{array}{c}
\sqrt{\frac{1}{2}+{\overline{\mathcal{S}}}_z} \\ 
\sqrt{\frac{1}{2}-{\overline{\mathcal{S}}}_z}\left(\frac{{\overline{\mathcal{S}}}_x}{\sqrt{(\frac{1}{2}+{\overline{\mathcal{S}}}_z)(\frac{1}{2}-{\overline{\mathcal{S}}}_z)}}+i\frac{{\overline{\mathcal{S}}}_y}{\sqrt{(\frac{1}{2}+{\overline{\mathcal{S}}}_z)(\frac{1}{2}-{\mathcal{S}}_z)}}\right) \end{array}
\right). 
\ee
i.e  ${\rho }_{\chi }=\left|{\chi }_0\right\rangle \langle {\chi }_0|$.

Our aim is to express  ${\overline{\mathcal{S}}}_x,{\overline{\mathcal{S}}}_y,{\overline{\mathcal{S}}}_z$ as functions by  ${\overline{s}}_x$,$\ {\overline{s}}_y$,$\ \overline{s_z}$,$\ {\overline{S}}_x,{\overline{S}}_y,{\overline{S}}_z$ and coefficients $c_1$and $c_2$.In order to have the expression of these averages ${\overline{\mathcal{S}}}_x,{\overline{\mathcal{S}}}_y,{\overline{\mathcal{S}}}_z$ as functions of arguments which are also the mean values {}{}of the spin projections we introduce the following relation of the complex numbers $c_1$,$c_2$ with the formal average meanings $0\le {\overline{\sum }}_1,{\overline{\sum }}_{2,}{\overline{\sum }}_3\le 1$ given by formulas for the projector $\left|{\psi }_c\right\rangle \left\langle {\psi }_c\right|$ where the spinor $|{\psi }_c\rangle $ reads:
\be
\left|{\psi }_c\right\rangle =\left( \begin{array}{c}
\sqrt{\frac{1}{2}+{\overline{\sum }}_3} \\ 
\sqrt{\frac{1}{2}-{\overline{\sum }}_3}\left(\frac{{\overline{\sum }}_1}{\sqrt{(\frac{1}{2}+{\overline{\sum }}_3)(\frac{1}{2}-{\overline{\sum }}_3)}}+i\frac{{\overline{\sum }}_2}{\sqrt{(\frac{1}{2}+{\overline{\sum }}_3)(\frac{1}{2}-{\overline{\sum }}_3)}}\right) \end{array}
\right).
\ee

We use the property that the phase of the complex coefficient $c_1$ can be chosen as zero as well as normalization condition ${\left|c_1\right|}^2+{\left|c_2\right|}^2$=1 is compatible with the final expression (27) of the density matrix of the superposition state.

Comparing the density matrix (27) with its expression (28) on terms ${\overline{\mathcal{S}}}_x,{\overline{\mathcal{S}}}_y,{\overline{\mathcal{S}}}_z$ we have the formulas:
\be
{\overline{\mathcal{S}}}_z=\frac{{\left|{c_1a}_1+c_2b_1\right|}^2}{{\left|{c_1a}_1+c_2b_1\right|}^2+{\left|{c_1a}_2+c_2b_2\right|}^2}.                                              
\end{equation} 
\begin{equation} \label{GrindEQ__29_} 
{\overline{\mathcal{S}}}_x\mathrm{-}\mathrm{i}{\overline{\mathcal{S}}}_y=\frac{\left({c_1a}_1+c_2b_1\right)\left({c^*_1a}^*_2+c^*_2b^*_2\right)}{{\left|{c_1a}_1+c_2b_1\right|}^2+{\left|{c_1a}_2+c_2b_2\right|}^2}.
\ee

Here we have to express the complex numbers $a_1,a_2,{b_1,b}_2,{c_1,c}_{2\ }$in terms of average meanings, i.e  $a_1=\sqrt{\frac{1}{2}+{\overline{s}}_z}$,$\ b_1=\sqrt{\frac{1}{2}+{\overline{S}}_z}$,$\ c_1=\sqrt{\frac{1}{2}+{\overline{\sum }}_3}$:
\be
a_2=\sqrt{\frac{1}{2}-{\overline{s}}_z}\left(\frac{{\overline{s}}_x}{\sqrt{(\frac{1}{2}+{\overline{s}}_z)(\frac{1}{2}-{\overline{s}}_z)}}+i\frac{{\overline{s}}_y}{\sqrt{(\frac{1}{2}+{\overline{s}}_z)(\frac{1}{2}-{\overline{s}}_z)}}\right).
\ee

\be
b_2=\sqrt{\frac{1}{2}-{\overline{S}}_z}\left(\frac{{\overline{S}}_x}{\sqrt{(\frac{1}{2}+{\overline{S}}_z)(\frac{1}{2}-{\overline{S}}_z)}}+i\frac{{\overline{S}}_y}{\sqrt{(\frac{1}{2}+{\overline{S}}_z)(\frac{1}{2}-{\overline{S}}_z)}}\right).
\ee

\be
c_2=\sqrt{\frac{1}{2}-{\overline{\sum }}_3}\left(\frac{{\overline{\sum }}_1}{\sqrt{(\frac{1}{2}+{\overline{\sum }}_3)(\frac{1}{2}-{\overline{\sum }}_3)}}+i\frac{{\overline{\sum }}_2}{\sqrt{(\frac{1}{2}+{\overline{\sum }}_3)(\frac{1}{2}-{\overline{\sum }}_3)}}\right).
\ee

The denominator in expression (32) reads:
\be 
T=1+\frac{2}{\sqrt{\left(\frac{1}{2}+{\overline{s}}_z\right)\left(\frac{1}{2}+{\overline{S}}_z\right)}}\left({\overline{\sum }}_1\left({\overline{S}}_x{\overline{s}}_x\mathrm{+}{\overline{S}}_y{\overline{s}}_y+\left(\frac{1}{2}+{\overline{s}}_z\right)\left(\frac{1}{2}+{\overline{S}}_z\right)\right)\ \ +{\overline{\sum }}_2\left({\overline{s}}_y{\overline{S}}_x\mathrm{-}{\overline{s}}_x{\overline{S}}_y\right)\right).
\ee

Finally we have:
\be
{\overline{\mathcal{S}}}_z=-\frac{1}{2}+\frac{1}{T}\left(\frac{1}{2}+{\overline{s}}_z+\left(\frac{1}{2}-{\overline{\sum }}_3\right)\left(\frac{1}{2}+{\overline{S}}_z\right)+2\sqrt{\left(\frac{1}{2}+{\overline{s}}_z\right)\left(\frac{1}{2}+{\overline{S}}_z\right)}{\overline{\sum }}_1\right).
\ee

\be
{\overline{\mathcal{S}}}_y=\frac{1}{T}\left({\overline{s}}_y\left(\frac{1}{2}+{\overline{\sum }}_3\right)\ +{\overline{S}}_y\left(\frac{1}{2}-{\overline{\sum }}_3\right)+\sqrt{\frac{\left(\frac{1}{2}+{\overline{S}}_z\right)}{\left(\frac{1}{2}+{\overline{s}}_z\right)}}\left({\overline{\sum }}_1{\overline{s}}_y\mathrm{-}{\overline{\sum }}_2{\overline{s}}_x\right)+\sqrt{\frac{\left(\frac{1}{2}+{\overline{s}}_z\right)}{\left(\frac{1}{2}+{\overline{S}}_z\right)}}\left({\overline{\sum }}_2{\overline{S}}_x\mathrm{+}{\overline{\sum }}_1{\overline{S}}_y\right)\right).
\ee

\be
{\overline{\mathcal{S}}}_x=\frac{1}{T}\left({\overline{s}}_x\left(\frac{1}{2}+{\overline{\sum }}_3\right)\ +{\overline{S}}_x\left(\frac{1}{2}-{\overline{\sum }}_3\right)+\sqrt{\frac{\left(\frac{1}{2}+{\overline{S}}_z\right)}{\left(\frac{1}{2}+{\overline{s}}_z\right)}}\left({\overline{\sum }}_1{\overline{s}}_x\mathrm{+}{\overline{\sum }}_2{\overline{s}}_y\right)+\sqrt{\frac{\left(\frac{1}{2}+{\overline{s}}_z\right)}{\left(\frac{1}{2}+{\overline{S}}_z\right)}}\left({\overline{\sum }}_1{\overline{S}}_x\mathrm{-}{\overline{\sum }}_2{\overline{S}}_y\right)\right).
\ee

\section{Conclusion}
\pst
We emphasize the main results of this work.

A new representation of the state of a particle with spin s = 1/2 was constructed, in which the state is identified with three average values of the spin projections m = + 1/2 into three perpendicular directions (on the x, y, z axes). For pure states, a new formulation of the principle of superposition was found in the form of a nonlinear rule of adding the average values of spin projections in these three directions. The formula for the addition of three projections of the spin was found explicitly, and the connection between this addition and the probabilistic representation of quantum states was determined by identifying states with probability distributions. This particle with spin $s =\frac{1}{2}$, found in [16-18]. The formulation of the principle of superposition of qubit states found in papers can be tested in experiments with particles with spin $s = \frac{1}{2}$. Subsequently, the resulting formula can be found for particles with higher spins, for example, for particles with spin $s=1$.

\section{Acknowledgements}
\pst
This work was supported by the Russian Science Foundation grant No.19-71-10091.


\begin{thebibliography}{99}

\bibitem{1}
E. Schrodinger,
{\it Quantisierung als Eigenwertproblem. Erste Mitteilung, Ann. Phys.}, \textbf{79}, 361 (1926).

\bibitem{2}
E. Schrodinger, {\it Naturwissenchaften},\textbf{ Bd. 14}, s. 664 (1926).

\bibitem{3} L.D. Landau, E.M. Lifshitz,
{\sl Quantum Mechanics ( Volume 3 of A Course of Theoretical Physics ) }Pergamon Press,1965.

\bibitem{4}
P. Dirac, {\it The Principles of Quantum Mechanics}, Oxford University Press (1930).


\bibitem{5}
A. S. Davydov, {\it Quantum Mechanics 2nd Edition}, 1st January (1965).

\bibitem{6}
Wigner E., {\it Phys. Rev.}, \textbf{40}, 749, (1932).

\bibitem{7}
Husimi K., {\it Proc. Phys. Math. Soc.}, Japan, \textbf{22}, 264, (1940).

\bibitem{8}
Glauber R. J., {\it Phys. Rev. Lett.}, \textbf{10}, 84, (1963).

\bibitem{9}
Sudarshan E. C. G., {\it Phys. Rev. Lett.}, \textbf{10}, 277, (1963).

\bibitem{10}
Glauber R. J.,Sudarshan E. C. G., {\it Fundamentals of Quantum Optics}, 428, (1970).

\bibitem{11}
Mancini S., Man’ko V. I.,Tombesi P., {\it J. Opt. B: Quantum Semiclass. Opt.},\textbf{7}, 615, (1995).

\bibitem{12}
Smithey D. T., Beck M., Raymer M. G., Faridani A., {\it Phys. Rev. Lett.},\textbf{70}, 1244, (1993).

\bibitem{13}
R.L. Stratonovich, Sov., {\it Phys. JETP},\textbf{4}, 891, (1957).

\bibitem{14}
S. Mancini, O. V. Man’ko, V. I. Man’ko, P. Tombesi, {\it The Pauli Equation For Probability Distributions},J. Phys. A: Math. Gen., 3461-3476, (2001).

\bibitem{15}
Vladimir I. Man'ko, Giuseppe Marmo, Franco Ventriglia, Patrizia Vitale, {\it J. Phys. A: Math. Theor. },\textbf{50}, 335302, (2017).

\bibitem{16}
V. N. Chernega, O. V. Man’ko, and V. I. Man’ko, {\it J. Russ. Laser Res. },\textbf{38}, 324, (2017).

\bibitem{17}
V. N. Chernega, O. V. Man’ko, and V. I. Man’ko, {\it J. Russ. Laser Res. },\textbf{38}, 141, (2017).

\bibitem{18}
V. N. Chernega, O. V. Man’ko, and V. I. Man’ko, {\it J. Russ. Laser Res. },\textbf{38}, 416, (2017).

\bibitem{19}
O. V. Man’ko, {\it Proceedings of Wigner Centennial Conference (Pecs, Hungary, 2002), The Official Electronic Proceedings, paper 30; Acta Physica Hungarica A, Series Heavy Ion Physics },\textbf{19}, 313, (2004).

\bibitem{20}
O.V. Man’ko, in: B. Gruber and M. Ramek (Eds.), {\it Proceedings of International Conference “Symmetries in Science X” (Bregenz, Austria,}, Plenum Press, New York, 207, (1998).

\bibitem{21}
V. V. Dodonov and V. I. Man’ko, {\it Phys. Lett. A}, \textbf{229}, 335, (1997).






\end{thebibliography}
\end{document}